# APPLYING MEMORY FORENSICS TO ROOTKIT DETECTION

**Igor Korkin**
National Research Nuclear University Moscow Engineering & Physics Institute (NRNU MEPhI)
Moscow, 115409, Russia
igor.korkin@gmail.com

**Ivan Nesterov**
Moscow Institute of Physics and Technology (MIPT)
Moscow Region 141700, Russia
i.nesterow@gmail.com

**Keywords**: Digital forensics, Virtual memory acquisition, Malware research, Rootkits detection, Anti-forensics.

## Content

## ABSTRACT

Volatile memory dump and its analysis is an essential part of digital forensics. Among a number of various software and hardware approaches for memory dumping there are authors who point out that some of these approaches are not resilient to various anti-forensic techniques, and others that require a reboot or are highly platform dependent. New resilient tools have certain disadvantages such as low speed or vulnerability to rootkits which directly manipulate kernel structures e.g. page tables. A new memory forensic system – Malware Analysis System for Hidden Knotty Anomalies (MASHKA) is described in this paper. It is resilient to popular anti-forensic techniques. The system can be used for doing a wide range of memory forensics tasks. This paper describes how to apply the system for research and detection of kernel mode rootkits and also presents analysis of the most popular anti-rootkit tools.

## 1. INTRODUCTION

Memory dump is used in various aspects of information security. It can be used for controlling virtual memory content while program is executed, running and after its close, is also typical for sophisticated malware, reverse-engineering due to it provides code and data in virtual memory for research and analysis. Memory dump is also used in computer forensic examination processes.

A fairly common problem is to obtain and analyze a memory dump. Both individual professionals J.Stuttgen, M.Cohen, B.Schatz, J.Okolica, J.Rutkowska, J.Butler, L.Cavallaro, L.Milkovic and entire international companies such as Microsoft, WindowsSCOPE, Guidance Software, Mandiant Corporation, Volatile Systems LLC tried to deal with this problem. A number of research theses are devoted to these issues [1-4].

It has also been discussed during various international conferences like BlackHat, DefCon, Digital Forensic Research Workgroup (DFRWS) Conference, ADFSL Conference on Digital Forensics, Security and Law, Open Source Digital Forensics Conference and workshops such as International Workshop on Digital Forensics (WSDF), SANS Windows Memory Forensics Training (FOR526), Open Memory Forensics Workshop (OMFW) by Volatile Systems.

This article presents a new memory dumping and analysis system which has several advantages and gives an example of how to use it for the kernel-mode rootkits and hidden malware detection. Moreover, this system can be applied in all mentioned above areas. The remainder of the paper is organized as follows.

Section 2 is devoted to the most popular software and hardware approaches for acquiring memory their analysis, including a new low-level approach. Memory dump can be obtained by executing a code that is running in user mode, kernel mode, VMX-root mode, system management mode and low-level AMT code which is used by an independent processor. These approaches can dump memory of single process address space or copy physical Random Access Memory (RAM). Tools and approaches focused on the mentioned code modes are described. As Microsoft Windows operating system is the most popular now it is essential to focus on OS Windows family of tools. However, similar conclusions could be made about Unix-based tools and approaches.

Section 3 contains a description of author's memory dump acquisition approach. The idea is based on walking through the page tables and saving each of them with additional information, such as virtual page addresses and its offsets in the result dump file. This approach reveals good efficiency when each page is not separately saved to HDD, but is buffered and archived before it is saved. Additional dump file encryption protects it from modification while it is being saved to HDD. This approach uses memory paging in protected mode and therefore is operating system independent and is applicable on Linux or Mac OS X.

In section 4 hidden malware is observed. The current available detection methods and tools are analyzed with the focus on signature detection of hidden drivers as the most common problem. An author's Dynamic Bit

Signature (DBS) and Rating Point Inspection (RPI) approaches for processes' and drivers' detection and comparative analysis are briefly presented.

Section 5 contains main conclusions and further research directions.

## 2. RELATED WORK

### 2.1. Virtual memory dump approaches

There are tools that can get a memory dump of the specified process, such as userdump.exe by Microsoft [5], pd.exe by T.Klein [6], pmdump.exe by A.Vidstrom [7], etc., which use OpenProcess and ReadProcessMemory functions or their low-level analogues like KeStackAttachProcess, ZwReadProcessMemory [8]. The review of these tools is outlined in the following papers [9], [10]. The first drawback of these approaches is their vulnerability to malware manipulation which can hinder expected behavior of these functions, for example by hooking them. The second drawback is that a corresponding dump file does not contain enough information for in-depth memory analysis. Some workarounds to solve these problems are presented further in this article.

### 2.2. Physical memory dump approaches

#### 2.2.1. Kernel Mode Code

Physical memory dump can be obtained on different levels of execution. There are three popular ways to obtain the dump in kernel mode: ZwOpenSection with ZwMapViewOfSection, MmMapIoSpace and MmMapMemoryDumpMdl.

Based on recently published papers and author's own reverse engineering research the internal mechanisms of some common commercial and free memory dump tools have been studied (see Table 1 for the listing of examined tools)

Table 1. Commercial and free memory dump tools

| Tool's title and version | Author |
|---|---|
| AccessData FTK Imager v.3.1.2.0 20. [11] | AccessData Group |
| Belkasoft Live RAM Capturer [12] | Belkasoft |
| Compiled Memory Analysis Tool (CMAT) [13] | J. Okolica, G.Peterson |
| DumpIt v2.0.0.20130807 RC1 [14] | MoonSols Ltd |
| Encase Forensic v.7.05 [15] | Guidance Software |
| FastDump v2.0.6.9 [16] | HBGary |
| Memory DD v1.3 [17] | ManTech International |
| Memoryze v3.0.0 [18] | Mandiant Corporation |
| ProDiscover Basic Edition v8.2.0.2 [19] | Technology PathWays |
| Redline v1.11 [18] | Mandiant Corporation |
| Winpmem v1.4.1 [20] | The Volatility Foundation |

It turns out that all these tools use one or several functions described above. Table 2 presents the results of the survey. Functions that are used in the program are marked with symbols «+» and «–».

Unfortunately KnTDD toolset by GMG Systems Inc [21] was unable to be obtained, but according to [22, 23] this toolset also uses the same functions.

Memory dump can also be acquired and analyzed remotely [24, 25], these possibilities are already implemented in commercial products, e.g. [26]. Toolset's local agent reads physical memory using the above mentioned functions and then transfers data to the server.

Table 2. Program tools and their functions

| Tool's name | Memory dump window functions | | |
|---|---|---|---|
| | ZwOpenSection, ZwMapViewOfSection | MmMapIoSpace | MmMapMemoryDumpMdl |
| AccessData FTK Imager | + | – | – |
| Belkasoft Live RAM Capturer | – | – | + |
| CMAT | + | – | – |
| Dumpit | + | + | + |
| Encase Forensic | + | – | – |
| FastDump | + | + | – |
| Memory DD | + | – | – |
| Memoryze | + | + | – |
| ProDiscover | + | – | – |
| RedLine | + | + | – |
| Winpmem | + | + | – |

Similarly to virtual memory dumping approaches malware can prevent memory acquisition, for example by hooking these functions.

Another method to prevent memory acquisition was described by L.Milkovic [23], where the author suggested hooking functions which save memory pages to HDD or transfer them and manipulate with buffers content. As a result final memory dump file will not contain pages with hidden objects including processes, drivers or network ports.

This clearly shows that the existing kernel-mode tools are not resilient to sophisticated malware.

### 2.2.2. VMX-root Mode Code

Let's focus on low-level approaches for memory dump acquisition. With the help of hardware virtualization technology it becomes possible to execute a code (hypervisor) on a more privileged level (VMX-root mode) than operation system's level. Hypervisors can be used to acquire memory dump. This process is described in the following projects [27, 28, 29].

Unlike the previously mentioned approaches this one is resilient to the most popular malware tricks which prevent memory dump acquisition. At the same time this method only works on systems, which support hardware virtualization and only in case when a previously loaded hypervisor supports nested virtualization [30].

One disadvantage of this method is its vulnerability to the "Man-In-The-Middle" attack, because malware hypervisor can load itself sooner than a trusted one. With the help of Shadow Page Tables (AMD) and Extended Page Tables (Intel) malware hypervisor can hide memory areas [31]. As a result the trusted hypervisor cannot read certain memory pages [32].

Trusted Execution Technology (TXT) by Intel and Secure Extension Mode (SEM) by AMD provides mechanism for a trusted hypervisor loading by means of Trusted Platform Module (TPM) [33, 34]. Unfortunately these technologies are also vulnerable [35, 36, 37].

This approach can be resilient to "Man-In-The-Middle" attack if a legitimate hypervisor is loaded from BIOS. However this case is only possible in laboratory conditions, because the BIOS hypervisor is highly platform dependent and its adaptation requires additional research that involves difficulties.

### 2.2.3. System Management Mode Code

System Management Mode (SMM) is more privileged than VMX-root mode. SMM provides power management features and backward compatibility. SMM is partially documented and described in the following papers by K.Zmudzinski [38], S.Embleton [39]. Opportunities of SMM to acquire memory dump were described in the following papers [40, 41].

Practical applicability of this method is hindered by installing of SMM dispatcher in general motherboard [42]. Another disadvantage of this approach is the necessity of PC rebooting that is not always possible [30]. This approach can be applied in some older models of motherboards. Adapting this method to new computers requires serious and non-trivial research.

**2.2.4. Active Management Technology Code**

On computers supporting Active Management Technology (AMT), which is a part of Intel Management Engine (ME), another memory acquisition method can be implemented. AMT code is executed in additional process unit which is located either in the Northbridge or Southbridge. As a result this code is more privileged than VMX-root mode code or SMM code.

The following papers cover this mode from the information security point of view [43, 44]. Due to the fact that malware can be executed in this mode [42, 45], we can state that memory dumping can operate in this mode too.

Widespread use of this method in practice is hampered by the lack of comprehensive documentation on AMT and ME.

**2.2.5. Hardware Approaches**

F. Davies in [46] mentions that with I/O Memory Management Unit technology (IOMMU) by AMD and Virtualization Technology for Directed I/O (VT-d) by Intel software approaches to memory acquisition will show poor performance if compared with hardware approaches. Therefore let us focus on hardware approaches to memory dump [47].

Capabilities of DMA devices such as PCI (PCIe) were used in the following tools: Tribble PCI Card by B.Carrier and J.Grand [48, 49], Co-Pilot by Komoku and Microsoft [31, 50], CaptureGuard PCIeCard by WindowsScope [51], RAM Capture Tool by BBN Technologies [50]. Capabilities of FireWire bus to acquire RAM memory were described by A.Boileau [52]. The applicability of hardware interfaces USB, eSATA, DisplayPort, Thunderbolt and others for accessing physical memory is described by R.Breuk and A.Spruyt [53]. These devices have a similar structure and are hardware boards, which are connected to a PC and designed for memory forensics.

Standard equipment can also be used to memory dump acquisition. For instance, usage of Graphics Address Remapping Table (GART) is described by N.Lawson, D.Goldsmith and T.Ptacek [54]. Y.Bulygin designed DeepWatch for memory dump acquisition with the help of the Northbridge integrated controller [55, 56].

It is essential to point out that malware can prevent memory dump acquisition even by hardware approaches. For example, External Access Protection technology by AMD is able to shadow memory pages from peripherals [57]. J.Rutkowska describes how to hide memory areas from peripheral access by reprogramming the Northbridge controller. Modifications in address dispatch tables in the Northbridge controller can hide physical memory regions [58].

Despite the fact that hardware approaches are resistant to common ways of hidden malicious software, they are only applicable under laboratory conditions, because of applicability and replication inconvenience.

**2.2.6. Other Software Approaches**

Among other tools for memory dump acquisition another approach was suggested with emulation tools such as VmWare, Vbox and others [10, 22]. This approach is based on suspending the virtual machine [9]. As a result the virtual machine paging file will contain the required data (*.vmem file in VmWare case). Malware is able to detect such emulation tools and hamper their work [59, 60].

Memory areas can also be acquired with the help of common operating system tools. Papers [10, 22, 23, 61] describe how to use pagefile.sys, crash dump file, hyberfil.sys for memory dump acquisition.

Page file is used for temporary storage of memory pages. According to papers [62, 63] the pagefile.sys does not contain full memory dump. To restore its content this file has to be merged with RAM dump, which poses additional difficulties.

Crash dump file (memory.dmp) will be created after a Windows system is crushed. This file contains information concerning the event details which caused the system crash. Microsoft developed a way to generate this file artificially – CrashOnCtrlScroll [64]. The disadvantage is that the crash dump is created only after the system is crashed, which is inconvenient for commodity systems. Crash dump file also has some other disadvantages [63].

Windows OS family starting with Vista adds support for hibernation mode. It causes creation of a hibernation system file (hyberfil.sys) which contains data about a current state of the system. On the one hand this file includes memory pages, but on the other hand it can hardly be used in deep forensic analysis. S.Vomel and F.Freiling [22] with reference to Russinovich point out that hyberfil.sys cannot be used to restore full RAM because of the limited quantity and quality of the saved pages file, this drawback is mentioned in [63].

There are a number of research projects based on the idea of 'cold booting', a method by S.Johannes, C.Michael [30]. Freezing memory chips, their removal from the computer and placing them into another PC to analyze memory content was suggested by Halderman [65] et al. Despite the fact that this idea has been extensively tested by several authors, it is still far from commodity production. This fact undoubtedly can be considered as a drawback [22, 63].

Another proof-of-concept project is BodySnatcher by Schatz which suggested using alternative OS injection on the top of the existing OS [66]. The main disadvantage of BodySnatcher is its poor usability, other disadvantages are described in the papers [63] and [22].

The latest approach to acquire a physical memory dump was offered by J. Stuttgen and M.Cohen in 'Anti-Forensic Resilient Memory Acquisition' [30]. With the help of rewriting page frame number in page table entries they got access to the required physical page. Their approach is resilient to modern anti-forensic techniques like hooking, but it is rather slow and vulnerable to rootkits which directly manipulate kernel pages table.

### 2.3. Conclusion

The analysis shows that the existing approaches and tools of memory dump acquisition do not fully comply with the current requirements:

1. Approaches based on Windows OS functions are vulnerable to intruder's attacks. VMX, SMM, AMT and hardware methods are difficult to use in industrial environments. They are more suitable for a specialized laboratory with highly qualified experts. Other research projects approaches are difficult to apply in practice.
2. Due to the fact that some memory pages are stored in a paging file, RAM dump does not contain complete data. This is especially obvious for PCs with low RAM.
3. The raw physical memory dump is not suitable for extracting useful information because relationships between the virtual and physical address spaces are lost. To overcome this fact additional work has to be done [67], for example lookup of EPROCESS structures by Burdach [68] or KPCR structures by Zwang, Wang [69]. This work involves a lot of difficulties.

It is essential to develop new detection software, which is resilient to common rootkits tricks. This software should pose great opportunities for memory dump analysis and forensics usage.

### 3. NEW MEMORY DUMP APPROACH

### 3.1. Overview

Researchers generally analyze the memory of the target process or kernel mode memory. In this paper we will be focused on the process context. We can use its copy from Windows integrated process such as notepad.exe or run an additionally installed process. As a result we can do research of either user mode or kernel mode memory.

To acquire memory dump for the specified process we run it and then attach to its context. To achieve anti-hook protection we use own low-level analogues of the following functions ZwCreateProcess and KeAttachProcess. As a result malware hooks are unable to hamper the memory acquisition.

Our analysis system does not only allow us to search different binary and text templates, but also do in-deep memory analysis. An example of such analysis will be given later.

The proposed system includes various tools to solve a lot of different memory content analysis tasks for the target program. It helps to investigate and detect malware and rootkits, reverse-engineer processes, conduct forensic research etc.

## 3.2. Details

### 3.2.1. Basics

It is suggested to launch one of the common processes or choose an already running one to analyze kernel mode memory. One of the possible scenarios may be the following: run notepad.exe, attach to it and dump memory, detach from it and terminate. When low level protected analogues of functions ZwCreateProcess and KeAttachProcess were developed, they were based on [70, 71, 72].

As a result of memory dumping two files will be created: the first file with memory pages 'dump.log' and the second one 'struct.log' with information about page virtual addresses and their offsets in 'dump.log'. Additional information about structures addresses, which are necessary for analysis, for example, EPROCESS list, KDBG, KPCR and etc. are saved into separate files. Examples of these files for analysis will be discussed later.

During dumping the content of each valid memory page is saved into 'dump.log' after buffering. Additional data is saved into 'struct.log', which includes virtual addresses of the pages beginning and end, offsets in 'dump.log' up to the beginning of the copied page. With the help of 'struct.log' and 'dump.log' it is possible to read page content, which corresponds to known virtual addresses and vice versa. Handling of these file is described in section 3.2.3.

### 3.2.2. Memory Dump Approach

Figure 1 shows an example of saving page #3 in 32-bit mode. It is well known that each virtual memory page corresponds to a page or frame in RAM. The corresponding pages sizes match but their order is usually different.

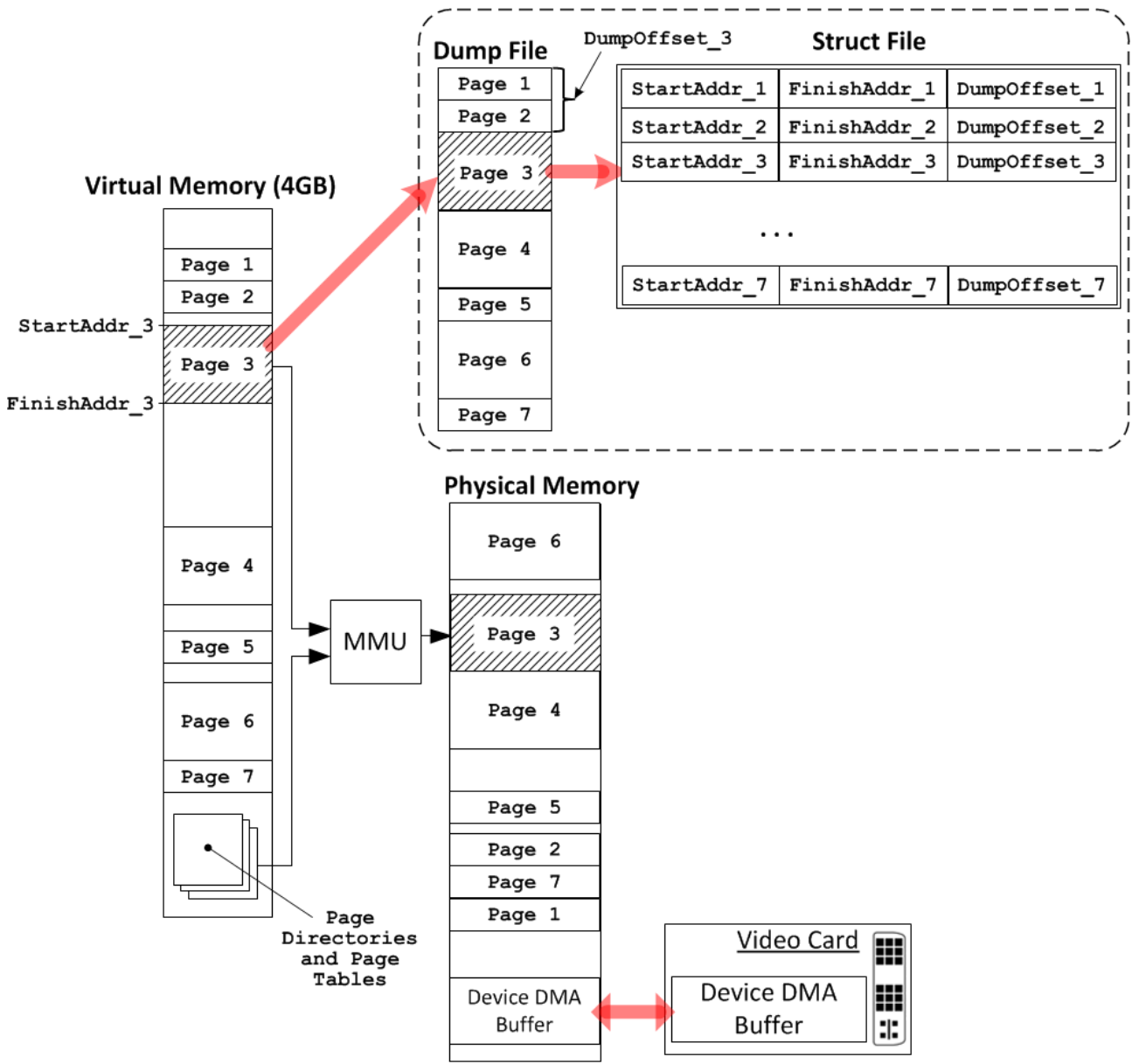

Figure 1. Memory dump acquisition process

Memory dump is acquired after walking through system tables such as Page Directories, Page Tables and others. Details of this walk depend on paging mode, whether Physical Address Extension (PAE) is enabled or not and also on 32-bit or 64-bit Windows versions.

Algorithms of walking for 32-bit Windows OS are similar whether PAE is enabled or not. In case of 64-bit the walking algorithm is similar but additional tables have to be taken into consideration. Therefore let us focus on the algorithm of walking for 32-bit Windows OS.

Memory dumping algorithm is based on the following tables walking workflow:

1. Walk successively through the Page Directory entries. Check the P flag of each entry.
2. If PDE.P is 0, go to the next entry. If PDE.P is 1, check the PS flag.
3. If PDE.PS is 1, save the corresponding memory page. Its size depends on whether PAE is enabled or not and is equal to 2-MByte or 4-MByte correspondingly.

4. If PDE.PS is 0, the current entry corresponds to the Page Table, which contains information about 4-KByte pages. Go to this Page Table.
5. In a similar way walk successively through the Page Table and check P flag of entry. If PTE.P is 0, go to the next entry, otherwise save the corresponding 4-KByte memory page.

Saving of each page is performed with buffering instead of getting copied directly into the file as it is done in a number of other tools. When the buffer is full, its content is being archived and encrypted and after that the results are saved into the 'dump.log'. Buffering helps to prevent these pages from modifying and increases the overall program performance [23].

Main features of the memory dump approach:

- The walk through the pages tables has to be done from high addresses up to low ones to exclude loading of empty pages. While walking from the first to the last entry CPU loads a lot of empty pages. Walking has to be started from the last entry to avoid this.
- The walk has to be implemented at PASSIVE_LEVEL IRQL, because only at this level accessing a page which is swapped to HDD means that its content is automatically loaded into memory.
- When we access a memory page related to device direct memory access (DMA) buffer system crash occurs in Windows Vista, 7 and 8. These critical exceptions cannot be caught by try and except. To prevent the crash these memory pages have to be ignored, see the details below.

Technique of ignoring memory pages of DMA devices

According to specification for PCIe (PCI) devices (for example modern network devices, video cards and others) they are able to directly access RAM. While we walk through virtual addresses OS functions allow getting physical addresses ranges of devices. To deal with this it is necessary to use Page Frame Number (PFN), which is a part of Page Table or Page Directory entries. The corresponding physical address is defined in the following way: PFN*0x1000. On the other hand virtual memory page address is determined with the help of indexes in Page Table and Page Directory.

To check whether this virtual memory page corresponds to the pages of DMA devices, the following steps have to be performed:

1. With the help of library functions exported from Setupapi.lib and Cfgmgr32.lib get the ranges of physical addresses which correspond to PCI devices ('prohibited list').
2. While walking through Page Table and Page Directory check each entry whether corresponding physical address belongs to 'prohibited list'. Once it does, skip this entry and check another one according to the algorithm.
3. If it does not, save the corresponding page according to the algorithm.

This technique has been successfully tested on several computers with different hardware and equipment. Access to the following PCI devices buffers (see Table 3) caused a system crash as described above.

Table 3. PCI devices which caused a system crash

| PC and OS | Devices which cause a system crash |
|---|---|
| Asus P5Q, Win7 32 | - NVIDIA GeForce GT 520;<br>- Atheros AR8121/AR8113/AR8114 PCI-E Ethernet Controller, integrated into motherboard Asus P5Q. |
| HP Z800, Win7 32 | - NVIDIA Quadro FX 580;<br>- D-Link DGE-560SX Single Fiber Gigabit Ethernet PCI-E Adapter (rev.A), additional plug-in device. |
| HP Z600, Win7 32 | |
| Shuttle XS36V, Win7 32 | No problem in basic configuration. |

The disadvantage of this method lies in ignoring physical memory ranges of all PCI devices to avoid crashes, no matter whether DMA is supported and used by this device or not. However it is possible to manually set the physical memory ranges that should be ignored. This disadvantage does not diminish the importance of MASHKA, because the essential structures such as EPROCESS and DRIVER_OBJECT cannot be located in the memory of these devices.

### 3.2.3. The Acquired Data Processing

Once 'dump.log', 'struct.log' and other files are received, they are processed either locally on a current PC or remotely after transferring these files to the remote host.

The main task of the dump analysis is gaining access to the dump data content located on the required virtual address. This operation is hampered in the existing products because there is not enough information about paging: whether virtual addresses correspond to physical addresses.

To achieve the correspondence between virtual addresses in original memory and offsets values in memory dump file we need additional two files - 'dump.log' and 'struct.log' simultaneously.

We will use the following abbreviations 'ODUF', 'VALF' and 'VAOM'. 'VALF' means the virtual addresses of the loaded memory dump file, 'ODUF' means corresponding offsets in dump file. File 'struct.log' contains virtual memory ranges of 'VAOM' and corresponding dump file offsets 'ODUF'. 'VAOM' is virtual address of the original memory; its values are used for further search for the structures, which contain the required virtual address.

Making memory dump analysis it is often necessary to use 'dump.log' and 'struct.log' files simultaneously and convert 'ODUF', 'VALF' and 'VAOM' into each other.

Let us look at this process.

1) 'VAOM' -> 'ODUF'
   As a result of the lookup in the 'struct.log' file we find i-entry, which contains virtual memory ranges, so that target value of 'VAOM' belongs to its range. 'ODUF' is defined in the following way: ODUF = DumpOffset[i] + (FinishAddr[i] - VAOM).
2) 'ODUF' -> 'VAOM'
   As a result of the lookup in the 'struct.log' file we find i-entry, so that Offset[i] <= ODUF < Offset[i+1], where Offset[i+1] means Offset of the following (i+1)-entry. 'VAOM' is defined in the following way: VAOM = FinishAddr[i] + (ODUF - Offset[i]).
3) 'ODUF' <-> 'VALF' and 'VAOM' <-> 'VALF'
   Values of 'ODUF' and 'VALF' are different by the value of starting address of the loaded dump file: VALF = ODUF + LoadAddr and vice versa. Having this equation it is possible to convert 'VAOM' <-> 'VALF'

These operations facilitate the in-depth analysis of the dump. Examples will be given below.

### 3.3. How to use MASHKA in memory forensics tasks

Memory analysis basic operations include text or binary signatures lookups through a memory dump. Current version of MASHKA can do multi-threaded lookup for the following objects: one byte (char) or wide-character (wchar_t) strings and byte fragments, for example addresses values.

When an object has been found, its VAOM, VALF and ODUF are forwarded for further research. The lookup is conducted from the beginning of the 'dump.log', byte-by-byte or in 4 byte order for special structures such as EPROCESS and DRIVER_OBJECT.

It is also possible to search the addresses, whose values are around the target VAOM address. For example, some system structures store information about the string values in the form of UNICODE_STRING or PUNICODE_STRING. To research and detect these structures it is necessary to conduct a search for the target wchar- string, and then further search for each discovered address of VAOM string. In case of PUNICODE_STRING it is necessary to conduct a search for the (VAOM-4) value, where 4 is 'Buffer' field offset from the beginning of UNICODE_STRING.

It is possible to search for byte fragment of target file header or one of its sections.

It is possible to carry out the following operations on the information obtained: walking through singly and doubly linked lists of structures and getting detailed information for further analysis, and also coping data located in the target virtual address range.

As an example, the following iterative research workflow for driver detection with the help of memory dump and WinDbg can be given:

1. Load OS Windows in debug mode under WinDbg control.
2. Install a driver with the specified 'ServiceName' and 'DisplayName' located in 'BinaryPathName'. Run this operation on the specified machine with the help of System Control Manager (SCM).
3. Hide this driver by well-known technique, based on PsLoadedModuleList [73].
4. Check the system with the help of some popular anti rootkit tools. This tool has to detect a deliberately hidden driver.
5. Get memory dump with the help of MASHKA. Copy 'dump.log', 'struct.log' and other essential files to the host machine.
6. Search for one byte and wide-character string containing, 'ServiceName', 'DisplayName' and 'BinaryPathName'. Save the received 'VAOM'.
7. Freeze the target machine with the help of WinDbg.
8. With the help of WinDbg change VAOM string content values. For example, patch one character 'A'-'Z' to the beginning of each string.
9. Check the system with the help of anti-rootkit for the second time. As a result the detection tools will give a changed name. Knowing the corresponding string names and 'VAOM' run further analysis. Sort out the details in the corresponding data lists, as well as in the anti-rootkits detection mechanisms.

## 4. NEW ROOTKITS DETECTION TOOL

This section is focused on the analysis of the existing approaches to hidden objects (processes and drivers) detection. Their drawbacks will be pointed out and author's detection approaches will be suggested, which uses Dynamic Bit Signature (DBS) for processes and Rating Point Inspection (RPI) for drivers. Finally, we will describe some currently known disadvantages of the approaches and ways to overcome them and for improvements.

### 4.1. Problem Statement

Cybercrime has become more and more sophisticated. Recently there has been a clear tendency or shift in computer attacks from mass infections to targeted attacks. E.Kaspersky assessed 'IT threats that have evolved from cyber hooliganism, via cybercrime to cyber warfare'. The new type of malware appeared such as Stuxnet, Duqu, Flamer, Gauss, that many antivirus companies call a cyber-weapon. Another example is spy network 'Red October' stole large amounts of data from diplomatic, government and science agencies in Europe, the Middle East and Central Asia for 5 years. Sophisticated intruder protection and heuristics did not prevent malware infection and subsequent activity [74].

Malware developers are working on long term attacks, which will give hackers an ongoing and virtually undetectable access to the target system [75, 76]. To ensure that malware has to use special rootkit mechanisms, which provide hiding of the following OS objects: processes, threads, drivers or services.

According to J.Rutkowska [77] there are two types of rootkit mechanisms to hide objects from built-in tools (for example 'taskmgr.exe' to get the processes list) which work in OS: functions-hooking mechanisms and direct kernel object manipulations (DKOM). Hooking is relatively simple to detect and will not be examined in this paper. Yet DKOM implementation uses minimal number of changes, which makes it the most complicated case for detection [78]. This case will be discussed later.

Current anti-rootkit approaches have significant disadvantages, i.e. they are either vulnerable, or their portability implies serious research.

Therefore the goal is to develop a new detection approach which is resilient to common rootkits tricks.

To develop a new detection method of rootkits let's examine how OS works. After the process has been executed, the OS creates a structure which solely corresponds to this process, see Figure 2.

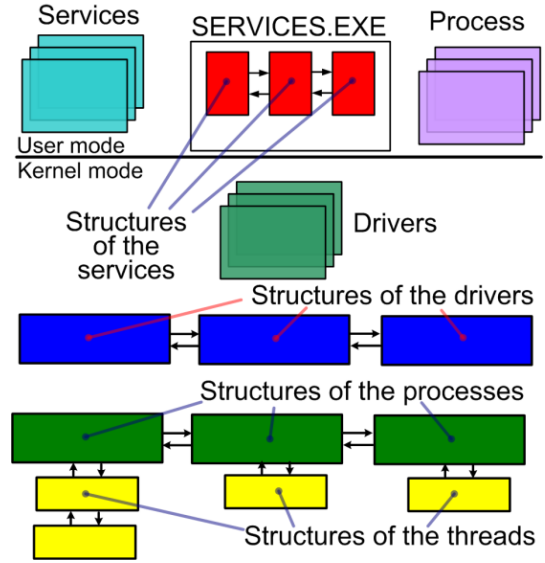

Figure 2. Processes, threads, drivers and services structures in Windows OS

Structures of different types are labeled as rectangles, for example, the process structures (EPROCESS) are labeled green. The object structures join OS doubly linked lists; see arrows on Figure 2. From these lists built-in tools get object information.

The idea of DKOM suggested by G.Hoglund was in concealing an OS object through the unlinking the corresponding structure. This does not crash the OS or yields to object termination, but built-in tools cannot further detect objects.

Next chapters describe existing methods and tools to detect hidden objects and their analysis.

### 4.2. Analysis of Current Approaches to Rootkit Detection in Face of Oppositions

One of the most popular ways to detect rootkits at runtime is known as cross-view detection, which relies on the fact that there are several ways to collect the same information about OS objects. Cross-view detection typically utilizes both high-level and low-level mechanisms to collect information [78]. The high-level mechanism is based on standard system functions to enumerate OS objects. The low-level mechanisms are based on data from some heuristic analyzers, additional object structure lists, signature scans and other heuristic [80].

We will analyze existing approaches according to a number of criteria, such as resilience to common rootkit tricks, portability to new versions of Windows and others.

Heuristic analyzer tracks programs activity, analyzes the collected data and blocks the program if its behavior is similar to a malicious one. The main disadvantage of this approach is that it blocks the program only after a certain amount of its activity has been collected during tracking. Another disadvantage is its vulnerability to rootkit countermeasures. Also heuristic analyzer must be started before malware, which is not always possible [81].

Information about running objects is often duplicated in different systems' lists. It is possible to use this data for objects detection. Is this case hidden object detection is based on data comparison obtained from various lists. This method was implemented in Tuluka Kernel Inspector, TDSS killer by Kaspersky lab and others. To hijack this detection the malware is able to modify all the needed lists to hide its own presence. As a result malware activity will not be detected.

Signature scan is based on the fact that values of some fields are either known or exceed the constant, for example 0x8000_0000. This method uses byte to byte search of fragments of objects structures in memory. This method has been implemented in GMER, PowerTool, XueTr and others. It is important to point out that structure sizes and their content change in new Windows versions (after some updates, service packs) as for EPROCESS structure. To deal with that, this method needs adaptation, which is often difficult because it requires manual adjustments.

It is possible to prevent hidden object detection by signature scan. To achieve this malware may modify some structure values, which are used by signature scan [73]. These modifications cannot crash the system or stop malware activity but make signature scan useless. One reason for this is that the decision is based only on the signature coincidence for the whole structure. If at least one byte does not match, the signature scan will miss the structure.

A similar method to prevent hidden object detection was proposed by T.Haruyama, H.Suzuki in 'One-byte Modification for Breaking Memory Forensic Analysis' article [82]. The prevention is based on modification of systems' structures values, which caused the situation when the detection tools were disabled.

Let us analyze the mentioned approaches with regard to processes structures (EPROCESS) and drivers structures (DRIVER_OBJECT) because they are often used in malware attacks.

### 4.2.1. Inside EPROCESS detection

When a process has been started a new content is created, and information about new object is added to different systems lists. A significant number of such lists make it difficult to hide the process well; therefore we usually speak about hiding the process only from built-in tools. There are a lot of approaches to process detection, so let us name some of them. There are some approaches based on additional objects structures lists, such as processes list from CSRSS.EXE, thread-based scheduling list and others [83, 84]. There are some heuristic analyzer approaches which are based on hooking functions, such as SwapContext or KiFastCallEntry. The Volatility Project includes various plugins list to stealth process detection [85, 86, 87].

Grizzard's approach [88] was based on locating x86 paging structures in memory images. Another MAS tool which was described in paper [88] uses memory crash dump file to rootkit detection, for this reason it is impossible to apply this method in commodity systems.

Another process detection approach has been suggested by Schuster [90]. This approach is based on the fact that values of some EPROCESS fields are either known or exceed the constant, for example 0x8000_0000. Author's approach has a number of important disadvantages: it is difficult to achieve its portability on different versions of Windows OS, as well as it is vulnerable to field modifications.

Another approach was based on signature search [91]. The authors suggest new graphs signatures, which can evaluate contingent structures in Linux OS. This method is also vulnerable to specific byte modifications. It is also difficult to make and test these graphs signatures for new Windows versions, because it requires a specialist's involvement.

Schuster's approach [90] was presented in paper [92]. It proposed including only robust fields in EPROCESS signature. If malware modifies one of these fields, the system crashes. To search these robust fields the author suggested control memory access with the help of adapted XEN hypervisor and VMware. The major drawback of this approach is its applicability only to structures with a lot of elements like EPROCESS, for which it is possible to find robust signature. Therefore it is impossible to apply this method to DRIVER_OBJECT structure detection.

### 4.2.2. Inside DRIVER_OBJECT detection

In comparison with process creation, driver loading causes much fewer system modifications, which makes it possible to achieve better drivers hiding.

Drivers hiding was described in popular books such as 'Rootkits: Subverting the Windows Kernel' [93] by G.Hoglund and J.Butler, and in new B. Blunden's book 'The Rootkit Arsenal: Escape and Evasion in the

Dark Corners of the System' [31]. It is necessary to mention drivers lists, which are not used by built-in tools: PsLoadModuleList [73], ObjectDirectory lists [73], Service Control Manager (SCM) drivers list [94].

Detection of hidden drivers is very similar to stealth process detection.

Schuster's signature approach [90] has been adapted by W.Tsaur and L.Yeh to drivers detection [95]. However, their approach is also vulnerable to target byte modification.

The following non-built-in well-known tools which support Windows 8 are: XueTr by linxer, PowerTool by ithurricane, TDSSKiller by Kaspersky Lab. In terms of driver detection three first tools have similar detection algorithms, which are based on byte-to-byte signature search among DRIVER_OBJECT structures. TDSSKiller uses a completely different detection algorithm. Its algorithm uses a system list, that holds information about new drivers added by SCM. By field values modifications it is possible to hide specified driver structures from all these tools. There are modifications that do not stop drivers or corrupt OS functionality.

It will be discussed further how to improve Schuster's idea [90] to create a rootkit detection approach, which is both resilient to byte modification and still portable to new Windows versions for both 32-bit and 64-bit editions.

### 4.3. New Stealth Processes and Drivers Detection Approach

#### 4.3.1. Dynamic Bit Signature (DBS) for EPROCESS Detection

Let us look first at the dynamic byte signature approach and how to apply it to process detection. After that we will describe dynamic bit signature approach and present its advantages.

To detect process structures hidden with the help of DKOM method we need to analyze the content of EPROCESS structures. Our goal is to find some common peculiarities between EPROCESS structures of different processes. Different bytes are illustrated on Figure 3 as squares with different colors. The corresponding squares have identical colors if the byte values are the same.

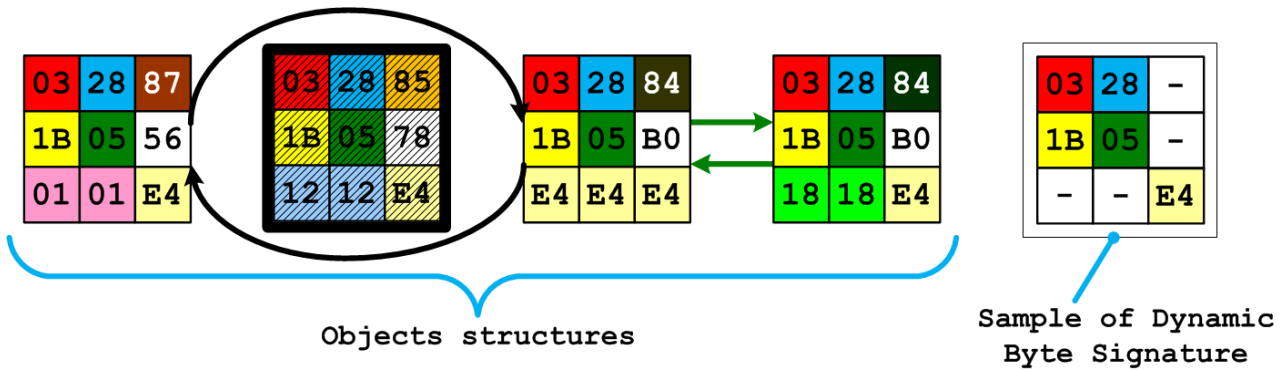

Figure 3. Objects structures typical design

It is obvious that initial bytes of each structure are identical, but further bytes are different. The conclusion was made, that if we search for some typical EPROCESS structure fragments it is possible to find all EPROCESS structures regardless of whether they are hidden or not. It is shown below how to do this.

Stealth process detection approach:

1. Create dynamic bit signature (DBS) as a template, which matches to all EPROCESS.
2. With the help of probabilistic search of DBS in kernel memory find all EPROCESS structures, either hidden or not. As a result, get the author's list.
3. Compare the author's list with a list of processes obtained by standard means of the OS, e.g. NtQuerySystemInformation

Dynamic bytes signature includes only the bytes, whose values are the same for all EPROCESS structures, which are in the list. For example, all EPROCESS structures contain the same byte in their center. It is labeled on Figure 3 as a green square ('05').

This byte is automatically added to the signature.

This signature in used to search EPROCESS structures manually. This is done with the help of byte-to-byte search in kernel memory. For each memory fragment the number of matches with DBS-signature is calculated. If for the current memory fragment the inequality $(\Sigma - \Delta) \leq i \leq \Sigma$ is true, it is considered that the structure of the similar object is found, $\Sigma$ – is the number of bytes in a signature, $\Delta$ – threshold value (for example $\Delta$ may be equal to $(\Sigma * 0.8)$), $i$ – is the number of matches for the current memory fragment with DBS-signature. If for some memory region this inequality is false, we skip this region and continue analysis with the next memory fragments until all the memory is analyzed.

As a result the full processes structures list based on DBS-signature matching will be obtained.

The conclusion if hidden processes are present is made after comparing DBS-matching list with the list obtained by NtQuerySystemInformation. This approach has been successfully tested for both cases of deliberately hidden objects and for real rootkits, such as Virus.Win32.Sality.q (Kaspersky Lab) and Trojan.Win32.VB.aqt (Kaspersky Lab).

It is important to emphasize that EPROCESS structure includes a lot of fields, whose values are linked with other kernel mode structures. Therefore these values exceed the values of 0x8000_0000. This fact is partly used in the Schuster's paper [90], but his approach is still vulnerable to byte modification and needs EPROCESS signature update when the new Windows version is released. We propose to improve the bytes-based signature approach with a bits-based one, which works in the similar way but on the bits values level.

Such approach has the following advantages:

- By the automatically generated bit-based signature, it is possible to adapt byte-based approach for new Windows versions and SP;
- Due to probabilistic nature of lookups it is possible to find all the EPROCESS structures even if they were deliberately modified and only 70-80% of data matches the signature. Threshold value can be adjusted manually.

This approach can be used to detect all objects in memory, which have a typical structure, but only if the structure definition is large enough. This method works badly for compact structures, because the amount of false detected structures increases. For detection of DRIVER_OBJECT structure, whose size is 4 times smaller than EPROCESS size, the proposed approach needs improvements that are described further.

### 4.3.2. Rating Point Inspection (RPI) for DRIVER_OBJECT detection

Rating Point Inspection (RPI) is the development of DBS detection approach. The first difference is that we need to manually adjust RPI to specific structure types such as DRIVER_OBJECT or DEVICE _OBJECT structures. The second difference in case on RPI is the utilization of additional weight matrix for precise matching accounting. We calculate total matching points (score) but not the individual matches themselves. For example, if one of the checks is true, 1, 2 or 3 etc. points are added to the final score. In DBS case we simply summarize the numbers of matches or add only 1 point to the final sum, if the check is true.

The conclusion for DRIVER_OBJECT structure matching is made in the similar way by comparing the score with the threshold value. The threshold value is determined by calculating the same metrics for "not hidden" DRIVER_OBJECT structures, which are located in DirectoryObject.

First let us briefly describe the DRIVER_OBJECT detection technique and then give an explanation:

1. Get memory dump ('dump.log' and 'struct.log'), save the DRIVER_OBJECT structures addresses in 'drvobj.log' file. To do the latter, use ZwOpenDirectoryObject function.

2. Determine 'min_major_function' value.
3. Determine 'global_scope' value.
4. Determine 'global_scope_deep' value.

The following steps (5,6,7) are done iteratively, and will be explained further.

5. Perform a byte-to-byte DRIVER_OBJECT structure search with the help of 'is_integrated_driver' function, which calculates the numbers of matching points for each memory region.
6. The conclusion that DRIVER_OBJECT structure is found is made after comparing these matching points from step 5 with the 'global_scope' value, which was obtained on step 3. If this value is not smaller than 'global_scope' value, the DRIVER_OBJECT structure is present. Otherwise calculate the numbers of deep matching points for this memory area with the help of 'is_integrated_driver_deep'. If the structure has been found, go to step 5 and continue lookups.
7. The conclusion that DRIVER_OBJECT structure is found is made after comparing the deep matching points obtained in step 6 with the 'global_scope_deep' value, which was obtained on step 4. If this value is not smaller than 'global_scope_deep' value, the DRIVER_OBJECT structure is present. Otherwise go to step 5 and continue lookups.
8. Repeat steps 6-8 for the whole memory area. As a result, get the RPI-matching list of DRIVER_OBJECT structures.
9. Compare the RPI-matching list with the drivers list, which has been obtained on step 1.

Further steps 2, 3 and 4 will be described in details further. Steps 6 and 7 are 'if-else' statements.

**Details of step 2. Determine 'min_major_function' value.** Use ZwOpenDirectoryObject function to obtain the list of DRIVER_OBJECT structures. For each DRIVER_OBJECT structure calculate the maximum number of functions' addresses from MajorFunction, whose addresses are the same, with the help of 'max_same_major_functions'. From these values select the minimum – 'min_major_function'.

**Details of step 3. Determine 'global_scope' value.** Use ZwOpenDirectoryObject function to obtain the list of DRIVER_OBJECT structures. For each DRIVER_OBJECT structure calculate the numbers of points with the help of Table 4. If one of the conditions is false, we add 0 points to the total number of matching points. Total matching score is calculated as a result of checking all the conditions in the table. For example, if all the conditions are true, apart from the second, the total score is 10. Among these values select the minimum – 'global_scope'.

Table 4. Weight matrix to calculate 'global_scope'

| Condition | Score |
| --- | --- |
| if (DRIVER_OBJECT_32.Type == 0x04) | 2 |
| if (DRIVER_OBJECT_32.Size == 0xa8) | 4 |
| if (chk_unicode_string(&DRIVER_OBJECT_32.DriverName)) | 2 |
| if (chk_unicode_string(DRIVER_OBJECT_32.HardwareDatabase)) | 2 |
| if ((DRIVER_OBJECT_32.MajorFunction[0]) >> 31) | 2 |
| if (max_same_major_functions(&DRIVER_OBJECT_32) >= min_major_function) | 2 |

Function '**chk_unicode_string**' checks whether the UNICODE_STRING structure is valid. This is done by checking conditions from the Table 5. Construction 'iswprint(UNICODE_STRING)' specifies checking of all the characters of the corresponding buffer using a 'iswprint' function.

Table 5. The 'chk_unicode_string' function

| Condition | Result |
|---|---|
| (UNICODE_STRING.MaximumLength >= UNICODE_STRING.Length) && (UNICODE_STRING.Buffer!=NULL) && iswprint(UNICODE_STRING) | true or false |

**Details of step 4. Determine 'global_scope_deep' value.** Use ZwOpenDirectoryObject function to obtain the list of DRIVER_OBJECT structures. For each DRIVER_OBJECT structure with the help of Table 6 calculate the numbers of matching points. Among these values select the minimum – 'global_scope_deep'.

Table 6. Weight matrix to calculate 'global_scope_deep'

| Condition | Score |
|---|---|
| if (DRIVER_OBJECT_32.Type == 0x04) | 2 |
| if (DRIVER_OBJECT_32.Size == 0xa8) | 2 |
| if (DRIVER_OBJECT_32.DriverStart >> 31) | 2 |
| if (DRIVER_OBJECT_32.DriverStart % 0x1000 == 0) | 2 |
| if (DRIVER_OBJECT_32.DriverSize % 0x1000 == 0) | 2 |
| if (check_function_prologue(DRIVER_OBJECT_32.DriverStart)) | 4 |
| if (DRIVER_OBJECT_32.DriverExtension >> 31 ) | 2 |
| K = chk_unicode_string2(&DRIVER_OBJECT_32.DriverName) | K |
| chk_unicode_string(DRIVER_OBJECT_32.HardwareDatabase) | 2 |
| if ((DRIVER_OBJECT_32.MajorFunction[0]) >> 31) | 2 |
| if (max_same_major_functions(&DRIVER_OBJECT_32) >= min_major_function) | 2 |

The function 'check_function_prologue' checks whether the conditions from the Table 7 are true. This check is repeated for first 16 memory bytes of each memory region (for (int i = 0; i < 0x10 ; i++)).

Table 7. The 'check_function_prologue(addr)' function

| Condition | Result |
|---|---|
| If (((addr[i+0] == 0x55) && (addr[i+1] == 0x89) && (addr[i+2] == 0xe5)) \|\| ((addr[i+0] == 0x55) && (addr[i+1] == 0x8b) && (addr[i+2] == 0xec)) \|\| ((addr[i+0] == 0x53) && (addr[i+1] == 0x56)) \|\| ((addr[i+0] == 0x56) && (addr[i+1] == 0x57)) \|\| ((addr[i+0] == 0x56) && (addr[i+1] == 0x57)) \|\| ((addr[i+0] == 0x8b) && (addr[i+1] == 0xff))) | true or false |

Function 'chk_unicode_string2' is determined in Table 8.

Table 8. The 'chk_unicode_string2(PUNICODE_STRING pDriverName)' function

| Condition | Score |
|---|---|
| if (pDriverName->MaximumLength >= pDriverName->Length) | 2 |
| if ((pDriverName->MaximumLength <= 0x50) && (pDriverName->Length <= 0x50)) | 4 |
| if (chk_unicode_string(pDriverName)) | 2 |
| if (_memicmp(pDriverName->Buffer, L".sys", pDriverName->MaximumLength)) | 2 |
| if (wcslen(pDriverName->Buffer) <= pDriverName->Length) | 2 |

RPI features and its further development:

It is possible to improve the function 'check_function_prologue' by adding an intelligent analyzer, which will detect modified function prologue. It is especially useful when malware employs any kind of armoring (e.g. packers, cryptors).

Also, it is possible for the detected hidden driver to look up its MD5 hash or name through Google search engine. Similar functionality has Process Explorer by M.Russinovich. It is well-known that sections contents on binary file in HDD or those which were loaded in memory do not differ much [96].

The RPI approach has been successfully tested for both cases of deliberately hidden objects, for real rootkits and for hidden drivers, which were loaded with the help of *ATSIV* utility by Linchpin Labs [97] and OSR. In the latter case all existing tools such as PowerTool, TDSSKiller, Xuetr cannot detect a hidden driver, but the proposed method can. YouTube video of these tools with comments is here [98].

In [95] it was proposed to utilize existing link between DRIVER_OBJECT and DEVICE_OBJECT structures to search for DRIVER_OBJECT structure. Unfortunately this link is optional and even conventional drivers structures may not have this relationship. It makes no sense to check this link. However, the RPI approach can be complemented by inspections of such links.

## 5. DISCUSSION AND FUTURE WORK

The presented MASHKA system has a number of advantages:

- Memory dump and analysis system, which is based on two shared files, have good opportunities for in-depth memory analysis and allow to find the hidden objects – processes and drivers. The first file contains pages contents and the second file contains corresponding sets of matches between virtual addresses and pages offsets.
- Protected implementation of memory dump avoids disruption from popular rootkits tricks.
- Bit-based signature approach provides the most profound inspection of system structures without manual work.
- Dynamic signature makes it possible to generate templates for byte-to-byte lookup or define signatures without a detailed study of the structure definition.
- Due to the fact that the matching conclusion is made with even partial matching to the signature, it is possible to detect even deliberately modified objects structures, where tools based on the idea of exact matching with the signature will miss the modified structure (e.g. Schuster's approach [89], GMER toolkit).

It is important to discuss how to use MASHKA to research and detect rootkits, which use modification of the page fault handler to hide memory pages, so called 'Shadow Walker'-like Rootkits. The bottleneck in MASHKA is linear search of structures templates, it is impossible to use GPU to increase its productivity. Logical development of this system is partial transition to the cloud – Anti Rootkit as a Service. The fact that vast majority of kernel mode structures are loaded into memory closely to each other was revealed. With the help of this fact it is possible to improve rootkit detection method. The cases of MASHKA application and implementation in education will be described later.

### 5.1. Detection Shadow Walker-like Rootkits

It is important to describe Shadow Walker rootkit (SW), which was presented by S.Sparks and J.Butler at the Black Hat conference in 2006 [99]. Despite the time passed this approach is still relevant. This rootkit can hide memory areas with the help of hooking the page fault interrupt handler. As a result, when accessing the memory pages containing the rootkit, their contents are replaced with false values.

Existing popular software [23, 30] does not detect rootkits of this type. Some authors propose to detect the rootkit using either program code, which works in more privileged mode than operation system (e.g. VMX mode or SMM), or hardware memory dump tools.

According to [100] this rootkit can be detected with the help of Interrupt Descriptor Table (IDT) analysis, because if SW has been installed, the page fault (#PF) handler is modified.

It is possible to detect this type of rootkits with also MASHKA. During the memory page walk we need to measure the duration of the memory page access. We need to make two successive attempts to access memory page. During the first access the memory page data loading occurs from page file to memory and system buffers (such as TLB) initialization occurs. The second memory access occurs when measuring the duration of memory page access. The memory region with too large access duration is the stealth memory region. Gaining access to the contents of this region depends on the rootkit implementation. For example it is possible to modify #PF handler. As a result, it is possible to control memory access and read hidden memory regions.

### 5.2. GPU Utilization in Memory Forensics

Detection of hidden objects occurs by memory lookups. Current version of MASHKA is based on C++ binary code with 'OpenMP' technology, which is provided by Microsoft Visual C++ compiler. However, the observed detection time can be significantly improved by utilizing Graphic Processing Unit (GPU), which is also occasionally called Visual Processing Unit (VPU). To do this we need to transfer the dump files to the device memory and perform all the algorithms on the GPU. The algorithms and memory lookups may be easily parallelized so that the analysis will speed up and CPU resources will be freed for common use.

### 5.3. The Idea of Cloud Anti Rootkit or Anti Rootkit as a Service

It is possible to use MASHKA toolkit system on tablet PC, such as ThinkPad Tablet, as well as on PC with low computational capabilities, such as low-cost laptops. The idea of cloud anti-rootkit or anti-rootkit as a service is as follows: data processing will occur remotely, not on the local PC. The separation of memory dumping and analysis processes yields to more reliable and more flexible IT security management infrastructure. More robust and solid dumping process may need very seldom updates but server-side application and algorithms need another maintenance periodicity. SaaS architecture simplifies the administration. The idea of cloud anti-rootkit leads to possibility of toolkit deployment in corporate networks without supplementary access to public Internet or with remote server in the cloud, so authorized users can load their memory dumps into the cloud and get the information whether there is any hidden object or not. While detecting hidden objects the system will provide detailed information and tools to analyze or eliminate these objects depending on usage scenarios.

### 5.4. The Center of Mass of Kernel Mode Structures

We have discovered another pattern which can be used in detection. Our research revealed that the placement of kernel mode structures such as EPROCESS and DRIVER_OBJECT are located closely to each other in memory. This fact can be used for detection of kernel mode structures. Based on the addresses of DRIVER_OBJECT structures the so-called 'center of mass' of DRIVER_OBJECT data can be found. The 'center of mass' will be located near most of the structures. When checking another memory area we need to assess how close it is to the 'centers of mass'. An additional criterion for detection is nearest to the 'center of mass' of the structure: the probability that the object found is the true structure increases as it approaches the 'center of mass'. We can calculate the 'center of mass' value with the help of addresses of kernel mode structures, which were already loaded in memory as a mean value.

This feature is valid for drivers loaded with the help of built-in mechanism, such as SCM. However, loaded by *ATSIV* utility by Linchpin Labs [97] this peculiarity is disrupted. To make it clear it is proposed to visualize a memory dump, reflecting the structures found. These issues are not covered in this paper.

### 5.5. Digital Forensics in Education

The proposed system can help students and postgraduate students in Computer Forensics to acquire practical skills. Students can get acquainted with the basics of memory forensics, Windows architecture, examine the program code and memory; investigate the relationships between binary modules loaded into memory. They will be able to learn the structure of user mode and kernel mode memories. The study of system services used to detect hidden objects during the training course may expect from the students to research the process SERVICES.EXE etc. Memory dump process evaluation makes it possible to study and get descriptions of undocumented structures of services that can be further used to search for hidden objects.

As a result, students consolidate their theoretical knowledge about the operating system, its components and their interaction with memory, as well as acquire research skills to get memory structures, which is crucial for solving practical problems of information security: reverse-engineering research and detection of malware, conducting forensic assessment and evaluation.

## ACKNOWLEDGEMENTS

We would like to thank Andrey Alexeevich Chechulin, research fellow of Laboratory of Computer Security Problems of the St. Petersburg Institute for Informatics and Automation of the Russian Academy of Science (Scientific advisor - prof. Igor Kotenko) for his insightful comments and feedback which help us to uplift the quality of the paper substantially.

## AUTHORS BIOGRAPHIES

| | |
|---|---|
| 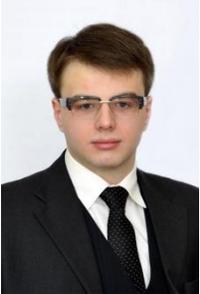 | Igor Korkin, Ph.D., is a specialist in information security. He works at Moscow Engineering & Physics Institute, training post-graduate students and supervising students. He has been engaged in rootkit technologies for over 6 years, he published more than 10 scientific papers. He was a finalist of the RusCrypto conference in 2011, with "Detection of nested virtual machine monitors" report, winner of "Hackers vs. Forensics" on Forum "Positive Hack Days 2012" in Moscow, Russia. He participated in a number of conferences and seminars. Research interests include: rootkits and anti-rootkits technologies; secure operating systems; spyware, backdoors and their detection; hardware virtualization; information leakage channels; memory forensics. |
| 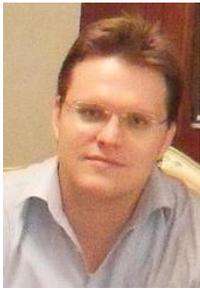 | Ivan Nesterov is an HPC Software specialist, system architect since 2000. His main research areas lie in the domain of high performance computing, parallel programming, distributed and storage systems, database design and applications. He finished the Moscow Institute of Physics and Technology (State University) with a M.Sc. in Applied Mathematics and Physics. Software design experience includes high-performance computing complex with hybrid CPU/GPU architecture for cryptography tasks, distributed visualization complex on heterogeneous computing systems with both non-uniform performance and architecture. Has GAZPROM IT Awards for best software application in 2009 and quality award in 2010. |